\documentclass [12pt]{article}
\usepackage{amsmath,amssymb,cite}
\usepackage{appendix}
\usepackage{feynmp}
\usepackage{float}
\usepackage[vcentermath,enableskew]{youngtab}
\usepackage{tabularx}
\usepackage[english]{babel}
\usepackage{graphicx}
\usepackage{indentfirst}
\usepackage{epsfig}
\usepackage{epstopdf}
\usepackage[]{hyperref}
\usepackage[section]{placeins}
\usepackage[stable]{footmisc}
\usepackage{appendix}
\usepackage{tabularx}
\usepackage[english]{babel}
\usepackage{graphicx}
\usepackage{indentfirst}
\usepackage{epsfig}
\usepackage{slashed}
\usepackage{fancyhdr}
\numberwithin{equation}{section}

\fancypagestyle{plain}{\fancyhead{}
}

\begin{document}

\title{
  Neutral monism, perspectivism and the quantum dualism: An essay
}

\author{Badis Ydri\\
Department of Physics, Annaba University,\\
 Annaba, Algeria.
}

\maketitle

\begin{abstract}
Quantum mechanics in the Wigner-von Neumann interpretation is presented. This is characterized by 1) a quantum dualism between matter and consciousness unified within an informational neutral monism, 2)  a quantum perspectivism which is extended to a complementarity between the Copenhagen  interpretation and the many-worlds formalism, 3) a psychophysical causal closure akin to Leibniz parallelism and 4) a quantum solipsism, i.e. a reality in which classical states are only potentially-existing until a conscious observation is made.

\end{abstract}

\tableofcontents

\section{Introduction} 

A revamping of the Wigner-von Neumann interpretation \cite{VN55,wigner} is presented which involves in an essential way i) perspectivism \cite{edwards} and quantum logic  \cite{bv} and  ii) naturalistic dualism \cite{chalmers} (quantum matter and classical consciousness are distinct aspects of a more neutral substance, i.e. neutral monism).

The resulting quantum dualism involves an extension of Bohr's complementarity \cite{Bohr} to a principle of complementarity between the local first-person subjective observers of the Copenhagen interpretation and the global third-person objective observers of the many-worlds formalism.

The collapse of the wave function (which is tied to the arrow of time and a comptabilistic account of free will) is seen as a psychophyscial force connecting the physical and mental similar (but acts in opposite direction) to the psychophysical force producing qualia. The relation between the physical and the mental is modeled on the many-minds interpretation \cite{AL88} producing therefore a mental causation similar to Leibniz parallelism (see \cite{leibniz} and references therein).

This form of causation produces also a weaker form of solipsism consistent with Bell's theorem \cite{Bell:1964kc,Bell:1964fg}. Indeed,  it is argued that this quantum dualism is characterized by an apparent subjective idealism (solipsism) in which the consciousness of the first-person observer causes the potentially-existing classical (pointer) states to actualize only in the sense of psychophysical parallelism.

In summary, this effective quantum dualism (with its fundamental description as an informational neutral monism) is quite different from both the classical Cartesian dualism \cite{descartes}and the classical Spinozian neutral monism \cite{spinoza}. It seems to avoid the two main problems of classical physicalism \cite{kim}:  i) the problem of mental causation (many-minds formulation and psychophysical parallelism) and ii) the hard problem of consciousness (naturalistic dualism). However, this single-world quantum dualism is dual (in virtue of quantum perspectivism) to a pure physicalism in a many-worlds where the Heisenberg cut \cite{Heisenberg} is placed at infinity, unitarity is the only law and there is no collapse.

\section{The collapse of the wave function and the measurement problem}

The Schrodinger’s cat experiment \cite{Schrodinger} is perhaps the most illuminating thought experiment which can be  used to delineate precisely what the measurement problem in quantum mechanics really is. The physical system under consideration consists of a radioactive atom plus a cat enclosed within the wall of a room together with a first-person observer called Schrodinger  standing outside the room.

A third-person observer (let us call him Wigner) is standing outside a larger room which encapsulates the Schrodinger's cat experiment, i.e. we are in fact considering a variant of the Wigner's friend experiment \cite{Wigner1} where Schrodinger plays effectively  the role of the friend.

A typical quantum process consists then of the following four stages:

\begin{enumerate}
\item The initial state of the observer outside the room is $|{\rm happy}\rangle$, of the cat inside the room is $|{\rm alive}\rangle$ and of the radioactive atom is $|{\rm undecayed}\rangle$.

\item After one hour there a fifty percent chance that the radioactive atom will decay thus activating a mechanism which releases a poison that kills the cat instantly. The states of the system radioactive atom+ cat after one hour is given therefore by the entangled state

\begin{eqnarray}
|\psi\rangle=\frac{1}{\sqrt{2}}(|0\rangle+|1\rangle). \label{form}
\end{eqnarray}
Where

\begin{eqnarray}
|0\rangle=|{\rm alive}\rangle|{\rm undecayed}\rangle~,~|1\rangle=|{\rm dead}\rangle|{\rm decayed}\rangle. 
\end{eqnarray}
The entangled state $|\psi\rangle$ means that a microscopic event (the decaying of the radioactive atom) is amplified to a macroscopic event (the life or death of the cat) thus the linear superposition principle which is known to be experimentally satisfied in all quantum situations is transferred to classical scales where only one branch of the wave function is observed to be realized at any one time. Indeed, the cat can be thought of as performing a measurement on the atom and finding it decayed and hence the cat dies in one branch of the wave function or finding it undecayed and hence the cat survives in the other branch of the wave function.

\item In the third stage the observer (Schrodinger)  makes his measurement on the state of the entangled system atom+cat by entering the room and looking at the cat thus becoming himself entangled with it. The complete entangled state is again given by a linear superposition of the form (\ref{form}), viz

\begin{eqnarray}
|\Psi\rangle=\frac{1}{\sqrt{2}}(|0\rangle+|1\rangle). \label{form1}
\end{eqnarray}
But the two branches $|0\rangle$ and $|1\rangle$ are now given by the states

\begin{eqnarray}
|0\rangle=|{\rm happy}\rangle|{\rm alive}\rangle|{\rm undecayed}\rangle~,~|1\rangle=|{\rm sad}\rangle|{\rm dead}\rangle|{\rm decayed}\rangle. 
\end{eqnarray}
Thus, in the branch $|0\rangle$ the observer is happy to see that the cat is alive because the atom did not decay whereas in the other branch $|1\rangle$ the observer is sad to see the cat is dead because the atom decayed. These two states are maximally entangled and that is why the state $|\Psi\rangle$ is called pre-measurement state. Indeed, this pre-measurement state contains coherence and interference between the branches and the pointer states $|0\rangle$ and $|1\rangle$ (which form the preferred-basis observed at the classical and macroscopic levels)  are still not actually realized. In fact, these states effectively do not exist before the completed measurement \cite{Bell:1964kc,Bell:1964fg}. The coherence and interference between the branches can be seen explicitly from the pure density matrix $\rho$ associated with the pure state $|\Psi\rangle$ given explicitly by

\begin{eqnarray}
\rho=|\Psi\rangle\langle\Psi|=\frac{1}{2}|0\rangle\langle 0|+ \frac{1}{2}|1\rangle\langle 1|+\frac{1}{2}|0\rangle\langle 1|+\frac{1}{2}|1\rangle\langle 0|.
\end{eqnarray}

The last two terms are the interference terms and interference, as poignantly stated by Feynman,  is the hallmark or even the mother of all quantum behavior.

From the perspective of the third-person observer (Wigner) the  conscious states (happy, sad) of the first-person observer (Schrodinger)  do not effectively exist before reducing the wave function (\ref{form1}) in the same way that the states (alive, dead) of the classical cat do not exist until the measurement is completed. Thus, the first-person observer in the pre-measurement state acts in some sense as a philosophical zombie. In the words of Wigner his friend Schrodinger is "in a state of suspended animation" before the measurement, i.e. while the superposed state $|\Psi\rangle$ given by (\ref{form1}) remains coherent.

\item The fourth and last stage is the completed measurement which is described by the reduced density matrix

\begin{eqnarray}
\rho_r=\frac{1}{2}|0\rangle\langle 0|+ \frac{1}{2}|1\rangle\langle 1|.
\end{eqnarray}
Thus, the off-diagonal elements responsible for interference are canceled and we end up with ordinary probabilities for mutual exclusive events, i.e. either the cat is alive or is dead with probability equal one half for each. However,  this is a mixed density matrix since there is no state $|\Psi\rangle$ in the Hilbert space for which $\rho_r=|\Psi\rangle\langle\Psi|$.

The completed measurement is therefore given by the (discontinuous, irreversible, instantaneous, non-deterministic and non-unitary) transition

\begin{eqnarray}
\rho\longrightarrow \rho_r.\label{reduction}
\end{eqnarray}
This is the collapse or reduction postulate and there is no known process in nature which effectuates this transition explicitly. This is the measurement problem. Decoherence \cite{zeh} tries to effectuate a collapse-like transition by means  of a unitary process which couples the system to the environment but this is beside the point since our fundamental working assumption here is that the above process (\ref{reduction}) is necessarily non-unitary with respect to the first-person observer.
\end{enumerate}

We have thus two fundamental quantum processes as formulated originally by von Neumann in his book \cite{VN55}:

\begin{itemize}
\item Process I which is given by the collapse of the wave function (occurring in stage four).

\item Process II which is given by the the unitary evolution in time generated by a Hamiltonian $H$, i.e. it is given by the Schrodinger equation (controlling stages one, two and three).
\end{itemize}

The Copenhagen interpretation is a broad interpretative framework of quantum mechanics due originally to Bohr \cite{Bohr} which can be mainly characterized by the assumption that the collapse of the wave function is a genuine independent and  fundamental process in nature (a fifth force of a sort) which is  not reducible to any of the other known interactions. This assumption is of course also shared by many other interpretations of quantum mechanics.

The weaker assumption that the collapse postulate is an approximation to some more fundamental unitary law of nature is the main characterization  of the many-worlds formalism \cite{mw2} and similar interpretations. See also \cite{manyworlds}.

These two standpoints are generally taken to be mutually exclusive interpretations of the laws of quantum mechanics which is quite an unfortunate situation. Indeed, it is always assumed either explicitly or implicitly that the collapse of the wave function and the subsequent actualization of the classical states of the cat as seen by the first-person observer (Schrodinger) is a fact in gross contradiction with the branching of the wave function, i.e. the real existence of a coherent linear superposition of the classical states of the cat as seen by the third-person observer (Wigner).

The central thesis of this essay is precisely the converse claim, i.e.  that the Copenhagen interpretation and the many-worlds formalism are not contradictory but they are in fact complementary. The Copenhagen interpretation provides the local perspective of reality  whereas the many-worlds provides the global perspective or view of reality.  This is similar to the black hole complementarity principle in which the descriptions given by the asymptotic (Schwarzschild) and infalling observers although they may look contradictory lassically are in fact complementary semi-classically.

Thus, a more complete and comprehensive interpretation of quantum mechanics admits the collapse of the wave function as a fundamental process of nature brought about by the observation of first-person local (or simulated as we will discuss in the conclusion) observers but also admits that the laws of nature are fully unitary, i.e. the branching of the wave function and the superposed states are real effect in the world with respect to  third-person global (unsimulated) observers. It  only seems that the world is full of the first kind of observers who exist in local system of coordinates not allowing them to access easily the full structure of the "curved manifold" of reality. See also Tegmark for example in \cite{Tegmark:1997me}.

An independent argument for the complementarity relation between the Copenhagen interpretation and the many-worlds formalism is given by the thought experiment considered  By Susskind in \cite{Susskind:2016jjb} which involves among other things the properties of entanglement entropy and black holes.

\section{Quantum perspectivism and quantum logic}

The observer plays a fundamental role in quantum mechanics.  The first-person observers of the Copenhagen interpretations (which dominate the world) provide the local view of reality while the third-person observers of the many-world formalism provide the global view of reality.

Therefore, quantum mechanics is strongly perspectival in character which was formalized using the language of quantum logic in \cite{edwards}.  In philosophy, perspectivism is the view due originally to Nietzsche  in which it is maintained that all reality is actually perspectival, i.e. there is no an objective reality out there independent from the knowing subject and free from interpretations and perspectives \cite{N0} (see also Leibniz and his theory of monads \cite{leibniz}). Perspectives for Nietzsche provide an "optics" of knowledge and they constitute the fundamental condition of the conscious observer in his search for value and meaning in existence and life. Somewhat more precisely, perspectivism can be interpreted as a middle position between metaphysical realism and relativism akin of Putnam's internal realism \cite{AL}.

For our purposes, every physical theory is characterized by a certain logic ${\cal L}$. For example, the logic of classical mechanics is a Boolean algebra ${\cal L}$ which is an orthocomplemented distributive lattice based on the power set of the phase space ${\Sigma}$. In other words, Classical events (also called experimental propositions) are subsets of the phase space, pure states semantically decide the truth value of experimental propositions, and the corresponding Boolean algebra underlies Kolmogorovian probability theory.

On the other hand,  it was shown by Birkhoff and von Neumann in their seminal paper \cite{bv} that the logic ${\cal L}$ underlying quantum mechanics is given  by the Hilbert lattice of projection operators on the Hilbert space ${\bf H}$ which is a non-Boolean, non-distributive and orthocomplemented lattice. Therefore the quantum events (or experimental propositions) are given in this case by closed linear subspaces of the Hilbert space, i.e. by projection  operators while pure states decides the truth value only probabilistically, and the corresponding Hilbert lattice or logic of projectors underlies the standard Born's rule.

The differences between the Boolean algebra of the phase space and the Hilbert lattice of projectors stems mostly from the logical disjunction operation ${\bf OR}$ which in the classical case is given by the union of subsets of the phase space whereas in the quantum case it is given by the direct sum of closed linear subspaces of the Hilbert space.

The problem of the classical limit in this context is seen as the problem of which projectors and their corresponding closed linear subspaces will tend in the limit $\hbar\longrightarrow 0$ to localized subsets of the phase space. This is clearly a highly non-trivial problem since the vast majority of projectors in the Hilbert space will certainly fail to admit any recognizable  classical limit.

In the classical case there is therefore a single perspective (corresponding to the Boolean structure of the classical logic) relative to which we can observe every possible measurable property of the system but in the quantum case there is no single privileged   perspective but instead there is an intricate web of non-trivially interlocked classical perspectives each of which corresponds to a maximal Boolean subalgebra (also called a block) of the Hilbert lattice. A block corresponds naturally to a maximal number of compatible and therefore comeasurable observables which  can only be defined locally but not globally. In other words, every block is only locally measurable, i.e. it can not be defined globally independently of the simultaneous measurement of the other blocks.

The Hilbert lattice in two dimensions can be viewed as a non-trivial disjoint union (pasting) of blocks. However, by Gleason's theorem \cite{gleason} and its generalization the Kochen-Specker theorem \cite{KS67} the Hilbert lattice does not admit in general a Boolean reduction, i.e. there is no a homomorphism from the Hilbert lattice into a Boolean algebra or to a disjoint collection (pasting)  of Boolean algebras which is perhaps the best characterization of the measurement problem. It is worth pointing out that the insistence on the Boolean character is precisely the assumption of reality envisaged by the EPR argument \cite{EPR35}.

Thus, in quantum mechanics the Hilbert lattice admits a decomposition into maximal Boolean subalgebras or blocks which define an intricate web of non-trivially interlocked classical perspectives. These perspectives are complementarity to each other and their totality defines an omni-perspective (which is seeing everything  from everywhere as defined by Nietzsche originally) which is the maximal possible perspective allowed by quantum mechanics. These perspectives are associated naturally with the first-person observers of the Copenhagen interpretation.

The perspective of the third-person observer of the many-worlds formalism which sees coherent linear superpositions of classical states (such as a dead and alive cats) is  a non-perspective  (which is seeing everything from nowhere, i.e. a God's eye view of a sort which is again Nietzsche's terminology) which is logically impossible according to  Nietzsche. However, this impossibility is simply due as we have seen to the fact that the world is full of first-order observers who are not directly aware of coeherent  linear superpositions, i.e. consciousness is through and through classical which is our second most important thesis in this essay.

Hence the complementarity between the many-worlds formalism and the Copenhagen interpetation proposed in this essay is nothing else but an extension of Bohr's complementarity principle which holds among first-person observers providing  classical perspectives on the world \cite{Bohr}.

\section{Physicalism, naturalistic‬‬ ‫‪dualism and neutral monism}

The Heisenberg cut \cite{Heisenberg} which demarcates the boundary between the observer (classical) and the observed (quantum) plays for the first-person Copenhagen observer a role similar to the role played for the asymptotic Schwarzschild observer by the event horizon (which separates the inside and outside of the black hole). Indeed, the placement of the cut although arbitrary it should be thought of as separating the accessible quantum degrees of freedom (associated with the observed physical system or with the entirety of the physical universe) from the inaccessible classical  degrees of freedom (those associated with the observer or more precisely with her consciousness) in the same way that the event horizon separates the accessible degrees of freedom (those outside the horizon) from the inaccessible degrees of freedom (those inside the horizon).  Thus, the fundamental quantum duality proposed in this essay between observer and observed is really a duality between a strictly quantum physical universe  and an exactly classical consciousness. This is a naturalistic dualism not necessarily a Cartesian one as we will now elucidate further.

The location of the Heisenberg cut is quite arbitrary under the hypothesis of physicalism which entails in particular  the assumptions i) of material monism and  ii) that the physical world is causally closed. Indeed, if there is nothing else but the material substance then there is no intrinsic difference between the observer and the observed and the world is strictly causally closed. As a consequence there can be no Heisenberg cut and no collapse of the wave function. This picture holds true in the  many-worlds formalism. The Heisenberg cut is therefore inexistent with respect to the hypothetical third-person observers in the same way that for the infalling observer who is freely falling into the black hole the  horizon is no special place in spacetime since she sees nothing special happening there (equivalence principle). These global third-person observers although they are hypothetical (since they are not directly observed in the world around us) their existence is fully logical (in contrast to Nietzsche view).

However, there is another complementary perspective associated with the realistic local first-person observers of the Copenhagen interpretation and with respect to whom i) there is indeed a fundamental distinction between the observer (classical system) and the observed (quantum system) and ii) the collapse of the wave function is a genuine physical effect which is actually verified experimentally. See quantum Zeno effect \cite{Misra:1976by,zeno_experiment}.

It can be argued  in this case, by following von Neumann, that the most natural placement of the Heisenberg cut is the interface between the consciousness of the observer, considered as a different substance,  and the physical brain which is a part of the material substance (see for example \cite{stapp} and references therein). The degrees of freedom associated with consciousness are thus constituted of a mental substance (a kind of dark energy which underlies mental phenomena) which is inaccessible to the first-person observers since the only degrees of freedom which can be physically accessed  in the usual way by any observer (local or global) must be part of the material substance.   Thus, these degrees of freedom are inaccessible since they lie effectively behind a horizon (the Heisenberg cut) and therefore they are analogous to the degrees of freedom  found behind the horizon of a black hole which are inaccessible to the asymptotic observer.

Physicalism, i.e. the view that everything is reduced to matter and that the world is causally closed  underlies all of physics including quantum mechanics.

But this idea of physicalism (or something near enough as advocated for from a physicalist point of view in \cite{kim}) is already challenged in the philosophy of mind by many philosophers such as Nagel \cite{nagel} and Chalmers \cite{chalmers}.

For example, according to Chalmers in his theory ‫‪of "naturalistic‬‬ ‫‪dualism"‬‬, both the usual degrees of freedom of matter and the degrees of freedom associated with consciousness are equally fundamental and can not be reduced to one another. In this case an appropriate ontology is really a double-aspect theory (such as the neutral monism of Spinoza) in which the material physical substance and the mental psychological substance are two aspects of a more neutral and a more fundamental substance which could perhaps be information, i.e. the "it from bit" of Wheeler \cite{wheeler} (see also Sayre \cite{sayre}). This is therefore a non-reductive theory which contains, beside the usual physical laws, psychophysical‬‬ ‫‪laws‬‬ (or further facts)  which determine how the mental arises from the physical. More precisely, subjective experience or qualitative consciousness ‫‪(or qualia) may‬‬ ‫‪arise‬‬ ‫‪from (caused by) ‫‪the‬‬ ‫‪physical ‫‪but‬‬ ‫‪it‬‬ ‫‪is‬‬ ‫‪not‬‬ ‫‪entailed‬‬ ‫‪by‬‬ ‫‪the‬‬ ‫‪physical, i.e. it is ontologically distinct and hence it can not be reduced to it.

Qualia is the so-called hard problem of consciousness which can not be explained functionally in contrast with the easy problems of consciousness which according to Chalmers can be explained functionally or neuronally (lower-level) or cognitively (higher level). This phenomenal consciousness (also called qualia) is what is associated with the subjective first-person experiences (like those of the Copenhagen observers) whereas functional consciousness or awareness is what is associated with the objective third-person experiences (like those of the many-world observers).

The role of the third-person observers of the many-worlds formalism seems also to be very similar to the role of the philosophical zombies in Chalmer's theory in the sense that they are both logically possible although they are not directly seen in nature which means in particular  that qualia, sentience, thought, value and perhaps even intentionality and alike (which involve a first-person experience) require further facts for their explanation.

Thus, there exists an explanatory gap as we go from the objective third-person level to the subjective  first-person level of consciousness ‬‬ which can only be accounted for by an effective theory of naturalistic dualism which should admit neutral monis as an underlying fundamental theory.

Physicalism is then challenged by the subjective experiences (which are tied to the existence of qualia such as color) of first-person conscious observers existing in this world. But physicalism  seems also to be challenged by the experiences of these first-person observers, considered within the Copenhagen interpretation, who can observe the collapse of the wave function (a fact tied to the conscious experience of time and comptabilist free will).

Indeed, in the same way that neuronal interactions in the brain (themselves due to the interaction of the brain with the physical system mediated by the environment) give rise to the subjective experience of color (the redness of the red for example) which is ontologically distinct from the brain itself, the interactions of the ontologically distinct degrees of freedom associated with the consciousness of the observer with the observed physical system (mediated through the brain and then the environment)  give rise to the so-called collapse of the wave function (dead or alive cat). It appears therefore that the collapse as a  (fifth) force works in the opposite direction of the psychological force producing qualia.

\section{Quantum dualism, Wigner's friend and solipsism}

By accepting consciousness as ontologically different substance with independent  degrees of freedom beside the usual physical particles and fields of the material substance we reach the inescapable conclusion that the Heisenberg cut between the classical conscious observer and the  quantum observed system  should be necessarily and naturally placed at the demarcation line between the physical brain and the non-physical mind. At this point, the Cophenhagen interpertation is essentially and effectively reduced to its logical limit which is the Wigner-von Neumann interpretation also called "the mind causes collapse" interpretation.

The complementarity between the Copenhagen interpretation and the many-worlds formalism implies therefore a disitinct dualism (quantum dualism) between consciousness (here in the form of the collapse of the wave function which is caused by the measurement of first-person observers) and matter (here in the form of the unitary branching of parallel worlds and their linear coherent superposition as seen by third-person observers).

This is different from Cartesian dualism, shares some common features with naturalistic dualism but  really it should be viewed as some form of neutral monism in which reality is constituted neither of consciousness nor of matter but of a more neutral substance with consciousness and matter being two different facets of it (not necessarily the only ones). The quantum dualism used by the first-person observers existing in this world should also be seen as complementary to the physicalism used by the hypothetical third-person observers of the many-world formalism (who are akin to the philosophical zombies of naturalistic dualism).

However, in naturalistic dualism the qualitative consciousness states (color and qualia in general) although ontologically independent of the physical they are produced by a physical system (the brain) but in the quantum dualism it is the classical physical states that are produced (brought from potentiality to actuality) by the ontologically independent consciousness of the first-person observers (through the collapse) which leads directly to the (far-fetched on the view of many) proposition of solipsism. The complementarity relation between the Copenhagen interpretation and the many-worlds formalism is thus translated into a complementarity relation between solipsism and many-worlds which should be viewed as dual properties.

Indeed, if by Bell's theorem the classical pointer or preferred states of a quantum system do not actually exist (but only exist in potentiality) until a quantum measurement is completed, and if by the Wigner-von Neumann interpretation a quantum measurement requires for its completion the free action of a conscious observer, then it follows naturally and logically that the degrees of freedom of the consciousness of the observer are responsible for bringing the classical states from potentiality into actuality which we see as collapse.

A somewhat explicit mechanism for the action of consciousness on matter can be given in analogy with the many-minds interpretation \cite{AL88} as follows.

The hypothetical third-person observer (Wigner) of the many-worlds sees directly the coherent linear superposition (\ref{form1}), i.e. the superposition of the decayed-atom/dead-cat/sad-observer and undecayed-atom/living-cat/happy-observer (where the "observer" here refers only to the physical brain states of the first-person observer).

By assuming an infinity of classical mind states, associated with the non-physical degrees of freedom of consciousness, evolving stochastically in time with a probability given by the Born's rule then it is observed that each mind state evolves in a stochastic way to either being attached to the decayed-atom/dead-cat/sad-observer branch or to the undecayed-atom/alive-cat/happy-observer branch. In other words, the first-person observer (Schrodinger) in the superposition (\ref{form1}) becomes fully conscious of the content of the classical state to which it is attached which appears therefore, from his perspective, as a collapse. This is the sense in which "consciousness causes collapse" which is ultimately due to the fact that  the classical mind or consciousness and the quantum matter both obey Born's rule which indicates that the underlying fundamental neutral substance also obeys the Born's rule. Also, it is not difficult to appreciate that this form of mental causation is nothing but the Leibnizian  psychophysical parallelism \cite{leibniz}.

Thus, from the perspective of the subjective experiences of the first-person observer it is the mental degrees of freedom that guide the time evolution of the physical (solipsism of the Wigner-von Neumann interpretation) whereas from the fact that quantum mechanics is more fundamental than classical mechanics it is seen that it is the physical degrees of freedom that guide the time evolution of the mental (third-person observer of the many-worlds). This duality between the two descriptions is of course due to the fact that both the classical mental and the quantum physical obey in this scheme the Born's rule. But is really quantum mechanics more fundamental than classical mechanics or are they really two independent descriptions of two ontologically distinct substances of nature?

An independent argument for "the mind causes collapse" interpretation is the largely underestimated or even underrated  Wigner's friend experiment. Indeed, before Wigner performs his measurement the state $|\Psi\rangle$  of the joint system Schrodinger+cat+atom is given by the maximally entangled pre-measurement state (\ref{form1}). Wigner can surely ask his friend Schrodinger whether or not he saw a dead cat and then inspect the system cat+atom. The probabilities according to the Born rule are as follows:

\begin{itemize}
\item There is a probability $1/2$ that Schrodinger will say "yes" and the system from then on behaves as if it is in the state $|1\rangle$ of a dead cat.

\item There is a probability $1/2$ that the friend will say "no" and the system from then on behaves as if it is in the state $|0\rangle$ of an alive cat.
\end{itemize}
In other words, it is for certain that the friend Schrodinger will say that he found a dead or alive cat, as the case may be, before Wigner asked him. This means in particular that in the reference frame (so to speak) of Schrodinger the state vector, even before Wigner's measurement, was already either $|1 \rangle$ or $|0\rangle$ and not their linear combination, which is in gross contradiction to the quantum mechanical rule (\ref{form1}) verified experimentally to a great accuracy.

This is not to say that Schrodinger's position is less reasonable  since quantum mechanics assumes him (in the reference frame of Wigner) to occupy the linear combination $|\Psi\rangle$ which  implies in a clear sense as Wigner puts it: "that my friend was in a state of suspended animation before he answered my question" \cite{Wigner1}. In other words, third-person observes (with respect to whom there is no collapse) really act as if they were philosophical zombies.

This experiment shows also among other things the non-tenability of objective collapse models (every measurement will produce a collapse for everybody) as opposed to the subjective-collapse models such as the Copenhagen interpretation (in which every observer is assigned a collapse in her own measurement only).

In summary, quantum mechanics enjoys four properties, which if taken at face value, makes the corresponding interpretation fall an easy prey to solipsism (or subjective idealism). These are:

\begin{enumerate}
\item Quantum mechanics is naturally a dualistic theory in which the dual relation between observer and observed is lifted to a dual relation between consciousness and matter considered as two different aspects of a more fundamental and more neutral substance.

\item Quantum mechanics is strongly perspectival in character, i.e. the role of the observer is irreducible and objective reality is nothing less or more than an omni-perspective which is a total coherent sum of all the perspectives of all the observers existing in this world.

\item The world is not causally closed (taking into account the physical alone) if observers and consciousness can not be eliminated.  Indeed, the collapse of the wave function requires the action of a conscious observer. In other words, consciousness acts as a kind of "dark energy" with causal influence on matter.

\item The (classical pointer) states of the physical system (or the world) exist only potentially and their existence becomes actualized or realized only when a measurement is performed on the physical system (or the world) in accordance with Bell's theorem. In other words,  a transition from potential to actual existence occurs only when the wave function collapses. The collapse of the wave function acts therefore as a completely independent  "fifth force" in nature connecting the mental to the physical similar to the psychophysical force producing qualia which connects the physical to the mental.
\end{enumerate}
As it turns out, most of the interpretations of quantum mechanics and most of the philosophy of (quantum) physics are in fact an attempt to explain away the above four straightforward properties of quantum mechanics. The only known exception to this rule is the he Wigner-von Neumann interpretation (also called "the mind causes collapse" interpretation) which can be defined precisely by the above four properties.

These properties amount effectively to a solipsism which is a reality in which only the mind of the knowing subject or the observer is objectively real and everything else is only an appearance, i.e. an idea in the mind of the observer with no thing-in-itself behind it.

However, the form of solipsism suggested by property 4 is much weaker than this since it states that the classical states which are actualized by the act of observation are actually actualized by an objectively real psychophysical action of a conscious observer (since measurement requires a conscious observer in this scheme) and hence the action (not only the consciousness) of the observer is also real within this interpretation which should extend to the entirety of the psychophysical world.

\section{Conclusion and related topics}

The Wigner-von Neumann interpretation of quantum mechanics was presented in this essay. It is argued that this interpretation, if taken as an effective theory, will involve four interwoven elements which are: 1) quantum dualism, 2) perspectivism as an extended complementarity principle, 3) psychophysical causal closure and 4) solipsism. But as a fundamental theory the Wigner-von Neumann interpretation depicts a reality of neutral monism which is complementary to the physicalism of the many-worlds.

Indeed, within this scheme we have from one hand the fact that in the single-world the interactions of the ontologically distinct degrees of freedom associated with the consciousness of the classical observer with the physical degrees of freedom of the quantum physical system give rise to the collapse of the wave function whereas from the hand we have  the complementarity relation between the Copenhagen interpretation and the many-worlds formalism.  This means in particular that the degrees of freedom associated with the consciousness of the first-person observer in the single-world are dual to the purely physical degrees of freedom associated with the many-worlds. As a consequence the quantum dualism employed by the first-person observers is a complementary description to the physicalism employed by the third-person observers.

The property of solipsism, which is seen as dual to the property of many-worlds, can be given a much weaker import by appealing to the many-minds interpretation and the underlying neutral monism.

Finally, we briefly comment on two intimately related topics.

\subsection{Simulation hypothesis}

By analogy with the simulation hypothesis \cite{Bostrom} the first-person observers of the Copenhagen interpretation (who see the collapse of the wave function) play the role of the simulated beings populating the simulation whereas the third-person observers of the many-worlds formalism (who are wholly unitary) play the role of the biological beings who are running the simulation.

Therefore, the simulated conscious first-person observers are like players in a giant virtual reality game and what they observe in the simulation is the rendering of the content of the simulated environment which appears to them as the the collapse of the wave function when reality is finally experienced.

For quantum systems the simulated reality is not computed until the observer or the player seeks the experience or observation. This is simply due to the high cost of the calculations and to the limited resources available to the simulator. This is then the statement that the electron is not out there until observed (Bell's theorem). 

But in classical systems  the numerical calculation is low cost and therefore the computer performs the calculation well before the observation or rendering time. So the moon is really out there even when we are not looking at it.

The above definition of physical reality as the connection between the rendering-of-the-simulated-environment to the player and the collapse-of-the-wave-function seen by the observer is based implicitly on the assumption of finite computation resources and requirement of low computational complexity \cite{COSW}.

Furthermore, if we are living in a simulated reality it is easy to imagine that the beings running the simulation has also computed the other parallel branches of the world. In fact this is only natural from the simulators point of view. These beings are  the third-person observers of the many-worlds formalism who observe  directly coherent linear superpositions and thus the corresponding global structure of reality  not only the local one associated with Copenhagen.

The simulation hypothesis provides therefore a vivid metaphor or visualization of the laws of quantum mechanics. But it can also be considered as a genuine metaphysical theory in its own right. In other words, we are indeed likely to be among the simulated beings rather than among the biological ones. This provides therefore a powerful starting point for a new interpretation of quantum mechanics.

\subsection{Time and free will}

The fact that there are two types of change in time within the Copenhagen and Wigner-von Neumann interpretations means in particular that there are two fundamental measures of time which are not necessarily the same. The unitary evolution of the Schrodinger equation corresponds precisely to the objective physical time. Whereas the collapse of the wave function in the measurement process (as seen by first-person observers such as Schrodinger) corresponds  to the subjective psychological time experienced by consciousness. It is the psychological time that has an arrow which is perspectival in character. These two times are not necessarily identical and they are expected to be unified (resulting in truely quantum time) within the framework of a new psychophysical theory of quantum mechanics (quantum gravity) in which a fine-grained unitary time evolution must replace the gross-grained Schrodinger time evolution which terminates dynamically (leaving the collapse in its place) each time a measurement is performed by a conscious observer.

The existence of two measures of time: i) the objective physical time (unitarity) and ii) the subjective psychological time (consciousness and collapse of the wave function) could be at the root cause of the problem of temporal non-locality in neuroscience discovered in experiments by Libet in 1983 \cite{libet}.  It was discovered  in these experiments that the readiness potential in the brain (which precedes voluntary movement) begins to rise in the brain before the conscious decision to move. By thinking about the conscious decision to move as marking the subjective psychological time associated with the collapse of the wave function whereas thinking about the rising of the readiness potential as marking the objective physical time we can see that the discrepancy found by Libet can be interpreted as a slight failure of synchronization between the physical and the mental, i.e. the psychophysical parallelism is not exact.  This is usually interpreted as indicating a  compatibilistic free will.


\begin{thebibliography}{99}

\bibitem{VN55}
  J.~von Neumann,
  ``Mathematical Foundations of Quantum Mechanics,''
  First edition (1955), Princeton University Press, translated from the 1932 German original by R. T. Beyer.


\bibitem{wigner}
Eugene P. Wigner,
``Symmetries and Reflections,''
 Indiana University Press, Bloomington, 1967.


 
\bibitem{edwards}
D.~A.~Edwards,
``The mathematical foundations of quantum mechanics,''
Synthese (1979) 42: 1. https://doi.org/10.1007/BF00413704.



\bibitem{bv}
  G.~Birkhoff, J.~ von Neumann,
  "The Logic of Quantum Mechanics,"
  Annals of Mathematics, 2nd Ser. 37 (4): 823–843. JSTOR 1968621.


  
  \bibitem{chalmers}
D.~Chalmers,
``The hard problem of consciousness,''
D.~Chalmers,
``Naturalistic dualism,''
 In M. Velmans, S. Schneider, “The blackwell companion to consciousness,” Blackwell Publishers Ltd, 2007.
See also his book "The character of consciousness", Oxford University Press (2010).


\bibitem{Bohr}
  N.~Bohr,
  ``Essays 1958-62 on Atomic Physics and Human Knowledge,'' Wiley, 1963.
    N.~Bohr,
  "Discussions with Einstein on Epistemological Problems in Atomic Physics".
  From Albert Einstein: Philosopher-Scientist (1949), publ. Cambridge University Press, 1949. Niels Bohr's report of conversations with Einstein.


  
\bibitem{AL88}
  D.~Albert, B.~Lower,
  ``Interpreting the Many Worlds Interpretation,''
  Synthese, 77, pp. 195–213, 1988.

\bibitem{N0}
  Nietzsche, Friedrich Wilhelm, Walter Arnold Kaufmann, and R.J. Hollingdale. The Will to Power. New York: Vintage Books, 1964. §481, p. 267.
  Friedrich Wilhelm Nietzsche and Walter Arnold Kaufmann. The Gay Science; With a Prelude in Rhymes and an Appendix of Songs. New York: Random House, 1974. Section 374, p. 336.)
  Friedrich Nietzsche, On the Genealogy of Morality: A Polemic. trans. Maudemarie Clarke and Alan J. Swenswen. Indianapolis: Hackett Publishing, 1998.)


  \bibitem{AL}
Anderson, R. Lanier,
“Truth and Objectivity in Perspectivism,” Synthese (1998) 115:1-32.
  
  \bibitem{leibniz}
M.~Kulstad and L.~Carlin, Laurence,
"Leibniz's Philosophy of Mind", The Stanford Encyclopedia of Philosophy (Winter 2013 Edition), Edward N. Zalta (ed.).


\bibitem{Bell:1964kc}
  J.~S.~Bell,
  ``On the Einstein-Podolsky-Rosen paradox,''
  Physics {\bf 1}, 195 (1964).

 

 

\bibitem{Bell:1964fg}
  J.~S.~Bell,
  ``On the Problem of Hidden Variables in Quantum Mechanics,''
  Rev.\ Mod.\ Phys.\  {\bf 38}, 447 (1966).


  
  \bibitem{descartes}
Rene Descartes,
 ``Meditationes de prima philosophia, “ 2nd edition, 1642.


  \bibitem{spinoza}
Benedict de Spinoza. 
``Ethica, “  1677. 
A translation called Ethics by Andrew Boyle,
revised by G. H. R. Parkinson, is published by Everyman’s Library, 1989, 1992.


  \bibitem{kim}
J.~Kim,
``Physicalism or something near enough,''
Princeton University Press (2005).


\bibitem{Schrodinger}
  J.D.~Trimmer,
  "The Present Situation in Quantum Mechanics: A Translation of Schrodinger's 'Cat Paradox' Paper,''
  Proceedings of the American Philosophical Society 124:5 (Oct. 10, 1980), 323-338.

 
\bibitem{Wigner1}
  E.~Wigner,
  "Remarks on the Mind Body Question,''
  In: Mehra J. (eds) Philosophical Reflections and Syntheses. The Collected Works of Eugene Paul Wigner (Part B Historical, Philosophical, and Soci  o-Political Papers), vol B/6. Springer, Berlin, Heidelberg.

\bibitem{zeh}
  H.~D.~Zeh,
  ``On the Interpretation of Measurement in Quantum Theory,''
  Foundation of Physics, vol. 1, pp. 69-76, (1970).

  
\bibitem{mw2}
  H.~Everett,
  ``Relative State Formulation of Quantum Mechanics,''
  Reviews of Modern Physics. 29: 454–462, 1957.



\bibitem{manyworlds} 
  H.~Everett, J.A.~Wheeler, B.S~DeWitt, L.N.~Cooper, D.~Van Vechten, N.~Graham, 
  "The Many-Worlds Interpretation of Quantum Mechanics,"
  B.~DeWitt, N.~Graham, (editors).
  Princeton Series in Physics (1973). Princeton, NJ: Princeton University Press. p. v. ISBN 0-691-08131-X.

\bibitem{Tegmark:1997me}
M.~Tegmark,
``The Interpretation of quantum mechanics: Many worlds or many words?,''
Fortsch. Phys. \textbf{46}, 855-862 (1998)
[arXiv:quant-ph/9709032 [quant-ph]].
  
  
\bibitem{Susskind:2016jjb}
  L.~Susskind,
  ``Copenhagen vs Everett, Teleportation, and ER=EPR,''
  Fortsch.\ Phys.\  {\bf 64}, no. 6-7, 551 (2016).
  [arXiv:1604.02589 [hep-th]].
  L.~Susskind,
  ``ER=EPR, GHZ, and the consistency of quantum measurements,''
  Fortsch.\ Phys.\  {\bf 64}, 72 (2016).
  [arXiv:1412.8483 [hep-th]].


\bibitem{Heisenberg}
  W.~Heisenberg,
  ``Is a deterministic completion of quantum mechanics possible?,''
Translation by E.~Crull and G. Bacciagaluppi,
https://halshs.archives-ouvertes.fr/halshs-00996315.



\bibitem{gleason}
  A.M.~Gleason,
  "Measures on the closed subspaces of a Hilbert space,"
  Journal of Mathematics and Mechanics, 6: 885–893 (1957).

     

  \bibitem{KS67}
    S.~Kochen, E.P.~Specker,
    "The problem of hidden variables in quantum mechanics",
    Journal of Mathematics and Mechanics 17 (1967), no. 1, 59-87, Reprinted in [Spe90, pp. 235-263].


    
  \bibitem{EPR35}
    A.~Einstein, B.~Podolsky, N.~Rosen,
    "Can quantum-mechanical description of physical reality be considered complete?",
    Physical Review 47 (1935),777-780, Reprinted in [WZ83, pages. 138-141].

    
\bibitem{Misra:1976by}
B.~Misra and E.~C.~G.~Sudarshan,
``The Zeno's Paradox in Quantum Theory,''
J. Math. Phys. \textbf{18}, 756 (1977)

\bibitem{zeno_experiment}
Y.S.~Patil, S.~Chakram, M.~Vengalattore,
"Quantum Control by Imaging : The Zeno effect in an ultracold lattice gas,"
Phys. Rev. Lett. 115, 140402 (2015),
arXiv:1411.2678 [cond-mat.quant-gas].


  \bibitem{stapp}
H.~Stapp,
``Quantum mechanical theories of consciousness,''
 In M. Velmans, S. Schneider, “The blackwell companion to consciousness,” Blackwell Publishers Ltd, 2007.
See also his book "Mind, Matter and Quantum Mechanics," Springer, The Frontiers Collection. ISBN 978-3-540-89653-1 (2009).



  \bibitem{nagel}
T.~Nagel,
``What is it like to be a bat,''
Philosophical Review. LXXXIII (4): 435–450. Oct 1974.



\bibitem{wheeler}
 J.~A.~ Wheeler,
”Information, physics, quantum: The search for links,”
In Zurek, Wojciech Hubert. ”Complexity, Entropy, and the Physics of Information”, Redwood City, California:
Addison-Wesley (1990). ISBN 978-0-201-51509-1. OCLC 21482771.


  \bibitem{sayre}
K.~M.~Sayre,
Cybernetics and the Philosophy of Mind, University of Notre Dame Press, 1976.


\bibitem{libet}
  B.~Libet,
  ``Mind Time: The temporal factor in consciousness,''
  Harvard University Press. ISBN 0-674-01320-4 (2004). 






\bibitem{Bostrom}
N.~Bostrom,
``Are you living in a computer simulation,''
Philosophical Quarterly. 53 (211):243-255.



\bibitem{COSW}
T.~Campbell, H.~Owhadi, J.~Sauvageau, D.~Watkinson,
``On testing the simulation theory,''
arXiv:1703.00058[quant-ph].

  
\end{thebibliography}
\end{document}